\documentclass[11pt,a4paper]{article}
\pdfoutput=1

\usepackage{epstopdf}
\usepackage{textcomp}
\usepackage[utf8]{inputenc} 
\usepackage{authblk}
\usepackage[percent]{overpic}
\usepackage{multirow}
\usepackage{rotating}
\usepackage{caption}
\usepackage{subcaption}
\usepackage[english]{babel}
\usepackage{pict2e}

\usepackage{jcappub}

\definecolor{darkgreen}{rgb}{0.1,0.6,0.1}

\title{Radiopurity of CaWO$_4$ Crystals for Direct Dark Matter Search with CRESST and EURECA}

\author[a]{A.~M\"{u}nster,}
\author[a]{M.~v.~Sivers,}
\author[b]{G.~Angloher,}
\author[b,c]{A.~Bento,}
\author[d]{C.~Bucci,}
\author[d]{L.~Canonica,}
\author[a,e]{A.~Erb,}
\author[a]{F.~v.~Feilitzsch,}
\author[d]{P.~Gorla,}
\author[a,f,g]{A.~G\"{u}tlein,}
\author[b]{D.~Hauff,}
\author[h]{J.~Jochum,}
\author[i]{H.~Kraus,}
\author[a]{J.-C.~Lanfranchi,}
\author[d]{M.~Laubenstein,}
\author[h]{J.~Loebell,}
\author[j,k]{Y.~Ortigoza,}
\author[b]{F.~Petricca,}
\author[a]{W.~Potzel,}
\author[b]{F.~Pr\"{o}bst,}
\author[j,k]{J.~Puimedon,}
\author[b]{F.~Reindl,}
\author[a]{S.~Roth,}
\author[h]{K.~Rottler,}
\author[h]{C.~Sailer,}
\author[d]{K.~Sch\"{a}ffner,}
\author[f,g]{J.~Schieck,}
\author[a]{S.~Scholl,}
\author[a]{S.~Sch\"{o}nert,}
\author[b]{W.~Seidel,}
\author[b]{L.~Stodolsky,}
\author[h]{C.~Strandhagen,}
\author[a,b]{R.~Strauss,}
\author[b]{A.~Tanzke,}
\author[h]{M.~Uffinger,}
\author[a]{A.~Ulrich,}
\author[h]{I.~Usherov,}
\author[a]{S.~Wawoczny,}
\author[a]{M.~Willers,}
\author[a,b]{M.~W\"{u}strich}
\author[a]{and A.~Z\"{o}ller}

\affiliation[a]{Physik-Department, Technische Universit\"{a}t M\"{u}nchen, D-85748 Garching, Germany}
\affiliation[b]{Max-Planck-Institut für Physik, D-80805 M\"{u}nchen, Germany}
\affiliation[c]{CIUC, Departamento de Fisica, Universidade de Coimbra, P3004 516 Coimbra, Portugal}
\affiliation[d]{INFN, Laboratori Nazionali del Gran Sasso, I-67010 Assergi, Italy}
\affiliation[e]{Walther-Meißner-Institute für Tieftemperaturforschung, D-85748 Garching, Germany}
\affiliation[f]{Institut f\"{u}r Hochenergiephysik der \"{O}sterreichischen Akademie der Wissenschaften, A-1050 Wien, Austria}
\affiliation[g]{Atominstitut, Technische Universit\"{a}t Wien, A-1020 Wien, Austria}
\affiliation[h]{Physikalisches Institut, Eberhard-Karls-Universit\"{a}t T\"{u}bingen, D-72076 T\"{u}bingen, Germany}
\affiliation[i]{Department of Physics, University of Oxford, Oxford OX1 3RH, United Kingdom}
\affiliation[j]{Grupo de F\'{i}sica Nuclear y Astropart\'{i}culas, Universidad de Zaragoza, ES-50009 Zaragoza, Spain}
\affiliation[k]{Laboratorio Subterráneo de Canfranc, Paseo de los Ayerbe s.n., ES-22880 Canfranc Estación, Huesca, Spain}

\emailAdd{andrea.muenster@ph.tum.de}
\emailAdd{msivers@ph.tum.de}

\abstract{
The direct dark matter search experiment CRESST uses scintillating CaWO$_4$ single crystals as targets for possible WIMP scatterings.
An intrinsic radioactive contamination of the crystals as low as possible is crucial for the sensitivity of the detectors.
In the past CaWO$_4$ crystals operated in CRESST were produced by institutes in Russia and the Ukraine. 
Since 2011 CaWO$_4$ crystals have also been grown at the crystal laboratory of the Technische Universität München (TUM) to better meet the requirements of CRESST and of the future tonne-scale multi-material experiment EURECA.
The radiopurity of the raw materials and of first TUM-grown crystals was measured by ultra-low background $\gamma$-spectrometry.
Two TUM-grown crystals were also operated as low-temperature detectors at a test setup in the Gran Sasso underground laboratory.
These measurements were used to determine the crystals' intrinsic $\alpha$-activities which were compared to those of crystals produced at other institutes.
The total $\alpha$-activities of TUM-grown crystals as low as $1.23\,\pm\,0.06\,\text{mBq/kg}$ were found to be significantly smaller than the activities of crystals grown at other institutes typically ranging between $\sim\,15\,\text{mBq/kg}$ and $\sim \, 35 \, \text{mBq/kg}$.} 

\keywords{dark matter detectors, dark matter experiments}

\notoc
\begin{document}                                 
\maketitle

\flushbottom

\section{Introduction}
\label{sec:introduction}

The CRESST (Cryogenic Rare Event Search with Superconducting Thermometers) experiment aims at the direct detection of WIMP\footnote{Weakly Interacting Massive Particle} dark matter using scintillating CaWO$_4$ crystals operated as low-temperature detectors \cite{angloher12}.
The EURECA (European Rare Event Calorimeter Array) project \cite{kraus09} is a future tonne-scale multi-material experiment which combines efforts of cryogenic dark matter searches in Europe (CRESST, EDELWEISS\footnote{Exp\'{e}rience pour DEtecter Les WIMPs En Site Souterrain}, ROSEBUD\footnote{Rare Objects SEarch with Bolometers UndergrounD}) and possibly also of SuperCDMS\footnote{Super Cryogenic Dark Matter Search} in the US.
\\
Because of the expected low event rate in experiments searching for dark matter it is crucial to achieve an efficient reduction of backgrounds originating from cosmic radiation and natural radioactivity in and around the detectors. 
Therefore, these experiments have to be located in underground laboratories and require several layers of additional shielding as well as a method of active background discrimination.
In CRESST such a background discrimination on an event-by-event basis is achieved by the simultaneous detection of phonons and scintillation light produced in the CaWO$_4$ target crystals by a particle interaction \cite{angloher12, meunier99}.
Compared to electron recoils (e$^-$/$\gamma$-events) $\alpha$-events produce only $\sim$~22\,\% of light \cite{angloher12}, nuclear recoils (n, WIMPs) even less (between $\sim 2$\,\% for W and $\sim 11$\,\% for O \cite{strauss14})  which leads to different bands of event classes in the light energy - phonon energy plane (see, e.g., \cite{TungstenAlpha}).
WIMP scatterings are expected in the nuclear recoil bands below 40\,keV \cite{angloher12}. 
\\
In the past the CaWO$_4$ crystals used in CRESST were provided by different institutes in Russia and the Ukraine (commercial crystals).
Since recently, CaWO$_4$ crystals are also produced within the collaboration using a dedicated Czochralski furnace at the crystal laboratory of the Technische Universit\"{a}t M\"{u}nchen (TUM) \cite{erb13}.
The aim of this effort was to have direct influence on the selection of the raw materials, the crystal growth and the after-growth treatment in order to improve radiopurity and scintillation properties as, e.g., the light output of the crystals.
Both - a lower radioactive contamination as well as a higher light output - increase the sensitivity of the detectors for WIMP detection.
\\
In this work we have investigated the radioactive contamination of the raw materials used for crystal growth at the TUM (CaCO$_3$ and WO$_3$ powders as well as CaWO$_4$ powder produced from them) and of first crystals grown from these powders using ultra-low background $\gamma$-spectrometry.
Furthermore, two CaWO$_4$ crystals grown at the TUM were operated as low-temperature detectors.
With these measurements the radiopurity of the crystals could be investigated by using intrinsic $\alpha$-decays, of which each single decay is observed, for an absolute $\alpha$-activity determination.
As these events are located in the MeV range of the $\alpha$-band they are clearly separated from all other event types, especially from the region of interest for dark matter search \cite{angloher12}.
However, several $\alpha$-decays are correlated to the important background of low-energy $\beta$-decays and $\gamma$-emissions (in the keV range) in the same decay chains.
An example is the isotope $^{227}$Ac with Q$_{\beta}$=44.8\,keV and excited states of 24.5\,keV and 9.3\,keV.
Its activity can be calculated from the $\alpha$-emitters $^{227}$Th and $^{223}$Ra that are in equilibrium with $^{227}$Ac.
The results of $\alpha$-activities of the two TUM-grown crystals are compared to those of commercial crystals using data from the previous CRESST run concluded in 2011 \cite{angloher12}.

\section{Experimental Aspects}
\label{sec:experimental}

\subsection{Samples}
\label{subsec:samples}
The raw materials used for crystal growth at the crystal laboratory of the TUM are high-purity powders of CaCO$_3$ and WO$_3$. 
Batches of WO$_3$ powder with a chemical purity of 4N8 were obtained from Alfa Aesar\footnote{Alfa Aesar GmbH \& Co KG (Karlsruhe, Germany)} (Sample AA) and MV Laboratories\footnote{MV Laboratories, Inc. (Frenchtown, NJ, USA)} (Sample MV). 
Several batches of CaCO$_3$ powder with a purity of 5N were obtained from the companies Alfa Aesar (Sample AA), MV Laboratories (Sample MV) and Ube Material\footnote{Ube Material Industries, Ltd (Tokyo, Japan)} (Sample UM). 
The materials from Alfa Aesar, which were the first to be available, were used to synthesize CaWO$_4$ powder by the following solid-state reaction at a temperature of $1200\,^\circ\text{C}$:
\begin{equation}
\text{CaCO}_3 + \text{WO}_3 \rightarrow \text{CaWO}_4 + \text{CO}_2
\label{eq:reaction}
\end{equation}
All TUM-grown crystals were produced from the CaWO$_4$ powder via the Czochralski method. 
A grown raw crystal (= ingot) has a typical mass of $\sim 800$\,g, a diameter of $\sim 45\,\text{mm}$ and a height of $\sim 130\,\text{mm}$.
The growth was carried out in Rh crucibles under 99\% Ar and 1\% O$_2$ atmosphere \cite{erb13} followed by an after-growth annealing under pure O$_2$ atmosphere at a temperature of $1450\,^\circ\text{C}$ \cite{sivers12}. \\
All crystals investigated were produced in separate runs. 
After each growth the crucible with the residual melt was refilled with new CaWO$_4$ powder as well as parts of former ingots not usable as detectors.\footnote{The crucible has to be refilled to the same amount for each growth process as the melt has to reach a certain level to start the growth process.}
None of the crystals was solely produced from already crystallized material.
The successfully grown crystals investigated in this work are named after their respective growth number (13~-~40) indicating the production order.
After growth numbers 19 and 31 the crucible was cleaned removing the residual melt and refilled.
Therefore, the crystals investigated were grown from a fresh melt as well as from a melt reused for several growth processes:
\newline
\begin{center}
TUM13\,$\xrightarrow{}$\,TUM16 $\xrightarrow[\text{\scriptsize{after growth 19}}]{\text{\scriptsize{crucible cleaned}}}$ TUM20\,$\xrightarrow{}$\,TUM22\,$\xrightarrow{}$\,TUM27 $\xrightarrow[\text{\scriptsize{after growth 31}}]{\text{\scriptsize{crucible cleaned}}}$ TUM40
\newline
\end{center}
The crystal samples TUM13, TUM16, TUM20, and TUM22 were cut from the ingots with the corresponding growth number. 
From the ingots of growth numbers 27 and 40 two absorbers for CRESST low-temperature detectors were produced. 
During production the upper and lower parts of the ingot as well as a surface layer of $\sim5\,\text{mm}$ were removed. 
The crystal TUM27 is of cylindrical shape ($\O=40\,\text{mm}$, $h=40\,\text{mm}$) with a mass of $\sim300\,\text{g}$ which is the standard geometry of the crystals currently used in CRESST. The crystal TUM40, on the other hand, is shaped as a square prism ($32\times32\times40\,\text{mm}^3$) with a mass of $\sim 250\,\text{g}$.\footnote{Two crystals of this alternative design are currently used in CRESST.}\
All other crystals investigated were obtained from GPI~RAS\footnote{General Physics Institute of the Russian Academy of Sciences (Moscow, Russia)} and SRC~"Carat"\footnote{Scientific Research Company "Carat" (Lviv, Ukraine)} where they were also grown by the Czochralski method. 
Some information concerning these crystals can be found in references \cite{danevich11, yakovyna04, yakovyna08}.\\
 
\subsection{Measurement Techniques}
\label{subsec:radiopurityInvest}
The samples of CaCO$_3$, WO$_3$ and CaWO$_4$ powders were screened with a HPGe detector at the Canfranc Underground Laboratory (LSC) with a maximum overburden of 2450 m.w.e. \cite{bettini12}. 
The radiopurity of the crystal samples that were produced at the TUM (TUM13, TUM16, TUM20, TUM22) was determined with HPGe detectors at the low background facility STELLA (SubTErranean Low Level Assay) \cite{Arpesella1996} of the Gran Sasso Underground Laboratory (LNGS) at a depth of 3800\,m.w.e. \cite{heusser06}. 
All HPGe detectors used (at LSC and LNGS) are surrounded by lead and copper to shield against ambient radioactivity and are equipped with a system to suppress radon.
The most sensitive detectors used at the LNGS can achieve sensitivities of the order of a few 10\,$\mu\text{Bq/kg}$ \cite{heusser06}.
\newline
The crystals TUM27 and TUM40 were operated for several weeks as low-temperature detectors in a test setup at the LNGS \cite{KSchaeffner_PHD}.
The results of these measurements were compared to those of the commercial cystals which were operated as low-temperature detectors in a dark matter run between 2009 and 2011 in the main CRESST setup \cite{angloher12} also located at the LNGS. 
Whereas the test setup is only shielded by a layer of low-activity lead \cite{KSchaeffner_PHD}, CRESST is surrounded by several layers of shielding including copper, lead, polyethylene, a radon box and an active muon veto \cite{angloher12}.
\\
Table~\ref{tab:SumDet} summarizes all powder samples and crystals investigated including the respective measurement location and technique used.
\begin{table}[t]
  \centering
  \begin{tabular}{ccc}\hline

    \small{detector location} & \small{measurement technique} & \small{samples investigated in this work} \\ \hline \hline
 
    \small{LSC (HPGe)} & \small{$\gamma$-spectrometry} & \small{CaCO$_3$, WO$_3$, CaWO$_4$ powders} \\ 
    \small{LNGS (HPGe)}& \small{$\gamma$-spectrometry} & \small{crystal samples TUM13/16/20/22}  \\ 
    \small{LNGS (test setup)} & \small{low-temperature det.} &  \small{crystals TUM27/40} \\
    \small{LNGS (CRESST main setup)} & \small{low-temperature det.} & \small{commercial crystals (run 2009-2011 \cite{angloher12})} \\ \hline	
   
  \end{tabular}
  \caption{Summary of all samples and crystals investigated in this work together with their respective measurement location at the LNGS (Laboratori Nazionali del Gran Sasso) or LSC (Laboratorio Subterráneo de Canfranc) and used technique. See main text for details.}
  \label{tab:SumDet}
\end{table}
\newline
$\alpha$-activities were investigated for all crystals operated as cryogenic detectors.
To obtain these $\alpha$-activities from the raw data different steps including energy calibration and data quality cuts are necessary.
An accurate energy calibration in the MeV range is only possible by interpolating the response of the detectors in between identified $\alpha$-lines.
A calibration at energies even higher than the identified $\alpha$-lines was achieved by a linear extrapolation.
\newline
In every detector investigated the total $\alpha$-activity including all events in the $\alpha$-band in the energy range between $1.5\,\text{MeV}$ (where the first $\alpha$-lines occur) and $\sim 7\,\text{MeV}$ (where the highest energies of single $\alpha$-lines from the natural decay chains can be found) was determined.
The activities of single $\alpha$-lines could be determined for three detectors.
For all other crystals statistics or energy resolution in the MeV range were not sufficient\footnote{The relative energy resolution in the $\alpha$-range can differ from the resolution at lower energies as the detectors are optimized for the energy range below 40\,keV \cite{angloher12} which is relevant for dark matter search.} or electronic artifacts occurred for some high-energetic detector pulses impeding the correct determination of the deposited energy.
All errors quoted for $\alpha$-activities in the following were calculated from statistical uncertainties.

\section{Results and Discussion}
\label{sec:results}

\subsection{$\gamma$-Spectrometry}
\label{subsec:results_gamma}
The $\gamma$-spectrometry results of the powders used for crystal growth are shown in tables~\ref{tab:WO3+CaCO3}~and~\ref{tab:CaWO4}.
For the WO$_3$ powders only upper limits on the activities could be derived while for two of the CaCO$_3$ samples a contamination with $^{226}$Ra of $26\,\pm\,6\,\text{mBq/kg}$ and $58\,\pm\,5\,\text{mBq/kg}$, respectively, was measured. 
This contamination can be explained by the similar chemical properties of Ca and Ra.
As sample MV shows only upper limits on all activities, the MV material was chosen to be used further on for crystal growth.
\newline
Table~\ref{tab:CaWO4} shows the results for the activity of the CaWO$_4$ powder synthesized at the TUM (see equation (\ref{eq:reaction})). 
The only contamination found is due to $^{226}$Ra which is in agreement with the contamination of the CaCO$_3$ powder used in the synthesis.
The small sample of CaWO$_4$ powder from reference \cite{danevich11}, which was purified in a special way, also shows a contamination with $^{226}$Ra (see table~\ref{tab:CaWO4}). \\
\begin{table}[t]
	\centering
	\begin{tabular}{lcc||ccc}
	\hline
	Material &  \multicolumn{2}{c||}{WO$_3$} & \multicolumn{3}{c}{CaCO$_3$}\\
	Sample & Sample AA & Sample MV & Sample AA & Sample MV & Sample UM\\
	Sample mass & 83\,g & 400\,g & 53\,g & 100\,g & 100\,g \\
	\hline
	\hline
	Isotope &  \multicolumn{4}{c}{Activity [mBq/kg]} \\
	\hline	
	$^{228}$Ra & $<\;23$ & $<2.1$ & $<27$ & $<12$ & $<11$\\
	$^{228}$Th & $<\;\;6$ & $<1.6$ & $<33$ & $<21$ & $<15$\\
	$^{238}$U  & $<400$ & $<68$ & $<260$ & $<180$ & $<110$\\
	$^{226}$Ra  & $<10$ & $<2.2$ & $26\pm6$ & $<21$ & $58\pm5$\\
	$^{40}$K & $<43$ & $<20$ & $<83$ & $<90$ & $<81$\\
	$^{137}$Cs & $<7$ & $<1.2$ & $<7$ & $<6$ & $<4$\\
	$^{60}$Co & $<2$ & $<0.68$ & $<6$ & $<4$ & $<2$\\
	$^{227}$Ac & $<36$ & $<5.6$ & $<31$ & $<23$ & $<17$\\
	\hline
	\end{tabular}

\caption{Activities of radionuclides determined by ultra-low background $\gamma$-spectrometry at the LSC for the WO$_3$ and CaCO$_3$ powders from different suppliers of raw materials. Upper limits are given with 95\% CL, uncertainties at 68\% CL.}
	\label{tab:WO3+CaCO3}
\end{table}
\begin{table}[t]
	\centering
	\begin{tabular}{lccccc}
	\hline
	Material &  \multicolumn{2}{c}{CaWO$_4$} \\
	Supplier & TUM & ref. \cite{danevich11} \\
	Sample mass & 57\,g  & 50\,g \\
	\hline
	\hline
	Isotope &  \multicolumn{4}{c}{Activity [mBq/kg]} \\
	\hline	
	$^{228}$Ra & $<17$ & $<19$\\
	$^{228}$Th & $<10$ & $<7$\\
	$^{238}$U & $<450$ & $<31$ \\
	$^{226}$Ra & $28\pm6$ & $19\pm5$ \\
	$^{40}$K & $<65$ & $<120$ \\
	$^{137}$Cs & $<5.5$ & -  \\
	$^{60}$Co& $<2.6$ & $<7$\\
	$^{227}$Ac& $<36$ & - \\
	\hline
	\end{tabular}

\caption{Activities of radionuclides determined by ultra-low background $\gamma$-spectrometry at the LSC for the CaWO$_4$ powder synthesized at the TUM. For comparison, the data of CaWO$_4$ powder from reference \cite{danevich11} is shown. Upper limits are given with 95\% CL, uncertainties at 68\% CL.}
	\label{tab:CaWO4}
\end{table}
Table~\ref{tab:crystals} summarizes the results for the activities of the CaWO$_4$ crystal samples produced at the TUM.
The values for the activity of $^{226}$Ra, especially the limit of $<1.6\,\text{mBq/kg}$ in case of the crystal TUM20, indicate in comparison to the much higher activity of $28\,\pm\,6\,\text{mBq/kg}$ in the CaWO$_4$ powder (see table~\ref{tab:CaWO4}) that $^{226}$Ra is rejected during crystal growth.
The segregation coefficient which is defined as ratio of the concentration of the contaminant in the solid (crystal) and the liquid phase (melt) is estimated to be s$_{\text{Ra}}<0.12$ (90\% CL). 
The segregation of Ra has also been observed in reference \cite{danevich11} and can be explained by the larger ionic radius of Ra$^{2+}$ which disfavors its accommodation at the site of the Ca$^{2+}$ ion.\\ 
The rejected Ra is accumulated in the melt which means that its concentration in the crystals will increase with increasing growth number if the residual melt is reused after each growth.
Therefore the crystals TUM13 and TUM16 produced with a high growth number without intermediate cleaning of the crucible show a measurable contamination with $^{226}$Ra. 
A similar behavior is expected for $^{228}$Th \cite{danevich11}.
\\
In case of $^{238}$U no clear conclusions can be drawn from these measurements. 
A measurable contamination with $^{238}$U was found in the crystals TUM16 and TUM22. However, within errors the values are compatible with the limits determined for the other crystals. 
Earlier work has shown that $^{238}$U is rejected by the crystal with an estimated segregation coefficient of $s_U\approx0.3$ \cite{danevich11}. 
As in case of $^{228}$Th and $^{226}$Ra one would, therefore, expect a larger contamination with $^{238}$U in the crystals TUM13 and TUM16 which is not observed. \\
The crystal TUM13 shows a contamination originating from $^{40}$K with a value of $32\,\pm\,9\,\text{mBq/kg}$ (table~\ref{tab:crystals}). 
As no such contamination was found for the crystal TUM16 produced at a later growth, it is likely that the presence of this nuclide is due to the handling of the crystal. 
A small contamination ($\sim1\,\text{mBq/kg}$) with $^{137}$Cs can be found in all crystals (see table~\ref{tab:crystals}). 
It cannot be excluded from the determined limits on the activities shown in table~\ref{tab:WO3+CaCO3} that $^{137}$Cs was already present in the raw materials.
The crystal from reference~\cite{danevich11} also included in table~\ref{tab:crystals} shows a heavy contamination with $^{210}$Pb.
\\
\begin{table}[t]
	\centering
	\begin{tabular}{lcccccc}
	\hline
	Crystal & TUM13 & TUM16 & TUM20 & TUM22 & ref. \cite{danevich11} & ref. \cite{zdesenko05}\\ 
	Sample mass  & 155\,g & 191\,g & 214\,g & 310\,g & 142\,g & 191\,g \\
	\hline
	\hline
	Isotope &  \multicolumn{6}{c}{Activity [mBq/kg]} \\
	\hline
	$^{228}$Ra &$<2.7$ & $<3.4$ &  $<1.4$ &  $<0.96$ & $<6$ & $0.7\pm0.1$ \\
	$^{228}$Th & $3\pm1$ & $4\pm1$ & $<1.8$ & $<1.2$ & $<7$ & $0.6\pm0.2$ \\
	$^{238}$U & $<110$ & $60\pm30$ &  $<47$ & $40\pm20$ & $<165$ & $14.0\pm0.5$\\
	$^{226}$Ra & $7\pm1$ & $13\pm1$ &  $<1.6$ & $<3.4$ & $4\pm2$ & $5.6\pm0.5$ \\
	$^{210}$Pb & - & - & - & - & $4740\pm570$ & $<430$ \\
	$^{235}$U & $<3.7$ & $<3.2$ &  $<2.9$ & $<1.1$ & - & $1.6\pm0.3$ \\
	$^{40}$K & $32\pm9$ & $<14$ &  $<8.7$ & $<13$ & $<36$ & $<12$ \\
	$^{137}$Cs & $0.9\pm0.4$ & $0.9\pm0.4$ & $1.6\pm0.5$ & $1.0\pm0.4$ & - & $<0.8$  \\
	$^{60}$Co &  $<1.9$ & $<0.38$ & $<0.3$ & $<0.22$ & $<4$ & -\\
	\hline
	\end{tabular}

	\caption{Activities of radionuclides determined for the crystals produced at the TUM by ultra-low background $\gamma$-spectrometry at the LNGS. The crystals are named after the consecutive growth number. After growth~number~19 the crucible was cleaned and refilled with fresh CaWO$_4$ powder. For comparison, the data of crystals from references \cite{danevich11,zdesenko05} are also shown. Upper limits from this work, reference \cite{danevich11} and reference \cite{zdesenko05} are given with 90\%, 95\% and 68\% CL, respectively. All uncertainties are given at 68\% CL.}
	\label{tab:crystals}
\end{table}
\subsection{$\alpha$-Activity Determination}
\label{subsec:results_alpha}

Table~\ref{tab:AlphaAct} shows for the crystals operated as low-temperature detectors the net live time  after cuts, the total number of $\alpha$-counts and the total $\alpha$-activity in the energy range between $1.5\,\text{MeV}$ and $\sim 7\,\text{MeV}$ without any corrections due to branching ratios or cascade decays (see later).
The first two entries of table~\ref{tab:AlphaAct} refer to the two TUM crystals, commercial crystals are listed in the second part.
For channels marked with an asterisk electronic artifacts were observed.
The corresponding systematic error on the determined activity could not be estimated.
\begin{table}[t]
  \centering
  \begin{tabular}{lcccr@{$\pm$}l}\hline

    Crystal & Supplier & Live time [h] & Counts & \multicolumn{2}{c}{$A_{\text{$\alpha$, total}}$ [mBq/kg]} \\ \hline \hline
 
 	TUM27 & TUM & 313.50 & 416 & 1.23\,&\,0.06 \\
	TUM40 & TUM & 301.48 & 834 & 3.07\,&\,0.11 \\ \hline \hline
    VK33 (Ch5)~*& GPI RAS& 6805.16 & $98\,076$ & 13.34\,&\,0.04 \\ 	
    Verena (Ch21) & GPI RAS & 6179.43 & $96\,384$ & 14.44\,&\,0.05  \\ 	
    Maja (Ch29)  & GPI RAS & 5434.78 & $628\,834$ & 107.13\,&\,0.14 \\ 	
    Sabine (Ch33)~*& GPI RAS & 5195.33 & $17\,860$ & 3.18\,&\,0.02 \\ 	
    Wibke (Ch43)~*& GPI RAS & 6490.85 & $231\,939$ & 33.09\,&\,0.07  \\ 
    K07 (Ch45)~*& SRC ``Carat'' & 6630.97 & $171\,383$ & 23.93\,&\,0.06 \\ 
    Daisy (Ch47)  & GPI RAS & 7207.86 & $23\,707$ & 3.05\,&\,0.02 \\ 
    Rita (Ch51)  & GPI RAS & 7422.62 & $115\,523$ & 14.41\,&\,0.04 \\ 
    Zora (Ch55)  & GPI RAS & 7680.44 & $306\,536$ & 36.95\,&\,0.07 \\  \hline
    
  \end{tabular}
  \caption{Final live times after cuts, number of $\alpha$-counts and total $\alpha$-activities in the energy range between $1.5\,\text{MeV}$ and $\sim 7\,\text{MeV}$ (without any corrections) of all crystals operated as low-temperature detectors. The two TUM crystals were measured at the test setup at the LNGS while the 9 crystals listed below were operated in the CRESST dark matter run between 2009 and 2011 \cite{angloher12}. For channels marked with an asterisk electronic artifacts occurred which might lead to an additional systematic error for the determined activities (not estimated here).}
  \label{tab:AlphaAct}
\end{table}
\newline
Since the TUM crystals show total $\alpha$-activities of $1.23\,\pm\,0.06\,\text{mBq/kg}$ and $3.07\,\pm\,0.11\,\text{mBq/kg}$, respectively, (see table~\ref{tab:AlphaAct}) they are in the range of the so far radiopurest commercial crystals Daisy and Sabine with activities of $3.05\,\pm\,0.02\,\text{mBq/kg}$ and $3.18\,\pm\,0.02\,\text{mBq/kg}$, respectively. 
The other crystals show activities between $\sim 15\,\text{mBq/kg}$ and $\sim 35\,\text{mBq/kg}$ with the exception of the crystal Maja with an activity of $107.13\,\pm\,0.14\,\text{mBq/kg}$.
Since both TUM crystals were produced after several growths and, therefore, not from a fresh melt, an even better radiopurity could be achieved by growing a crystal from a fresh melt.
The large spread in $\alpha$-activities of commercial crystals from the same institute results possibly from the use of different raw materials and differences in the growth process.\footnote{The crystal K07 shows - in comparison to the other commercial crystals - a high $^{210}$Pb contamination of $\sim20$\,mBq/kg.
This is, however, still two orders of magnitude smaller than the $^{210}$Pb activity found in the crystal of reference~\cite{danevich11} also grown at SRC ``Carat'' (see table~\ref{tab:crystals}).}
\newline
Although high statistics were achieved for the commercial crystals the activities of single $\alpha$-lines could only be determined for the crystals Rita, Daisy and Verena.
\renewcommand{\arraystretch}{1.15}
\begin{table}[t]
  \centering
  \begin{tabular}{|cc|c|c|c|} \hline
	 & & Rita & Daisy & Verena \\
    Isotope & Q$_{\alpha}$-value [keV] & A [mBq/kg] & A [mBq/kg] & A [mBq/kg] \\ \hline \hline
    $^{144}$Nd & 1905.2 & $< 0.002$  & $0.018 \pm 0.002$  & $< 0.002$ \\ \hline
    $^{152}$Gd & 2204.6 & $< 0.003$  & $0.012 \pm 0.001$ & $< 0.003$ \\ \hline
    $^{147}$Sm & 2310.5 & $0.120 \pm 0.004$  & $0.975 \pm 0.011$  & $0.119 \pm 0.004$ \\ \hline
    $^{180}$W & 2516.4 & $0.037 \pm 0.002$  & $0.038 \pm 0.002$ & $0.039 \pm 0.002$ \\ \hline
    \color{darkgreen} $^{232}$Th & 4082.8 & $0.284 \pm 0.006$ & $0.049 \pm 0.003$ & $0.324 \pm 0.007$ \\ \hline
    \color{blue} $^{238}$U & 4270 & $0.041 \pm 0.002$  & $0.042 \pm 0.002$ & $0.055 \pm 0.003$ \\ \hline
    \color{blue} $^{230}$Th & 4770.0 & $1.422 \pm 0.013$ & $0.090 \pm 0.003$ & $1.534 \pm 0.015$ \\ \hline
    \color{blue} $^{234}$U & 4858.5 & \multirow{2}{*}{$0.973 \pm 0.011$} & \multirow{2}{*}{$0.124 \pm 0.004$} & \multirow{2}{*}{$0.975 \pm 0.012$} \\ \cline{1-2}
    \color{blue} $^{226}$Ra & 4870.6 & & & \\ \hline
    \color{red} $^{227}$Ac & 5042.2 & $1.594 \pm 0.145$  & \multirow{2}{*}{$0.360 \pm 0.008$}  & $1.792 \pm 0.145$ \\ \cline{1-3} \cline{5-5}
    \color{red} $^{231}$Pa & 5149.9 & $0.222 \pm 0.005$ &  & $0.247 \pm 0.006$ \\ \hline
    \color{blue} $^{210}$Po ext & 5304.4 & $0.142 \pm 0.004$& $0.108 \pm 0.004$  & $0.148 \pm 0.005$ \\ \hline
    \color{blue} $^{210}$Po int & 5407.5 & $3.420 \pm 0.021$ & $0.144 \pm 0.004$ & $2.545 \pm 0.020$ \\ \hline
    \color{darkgreen} $^{228}$Th & 5520.1 & $0.556 \pm 0.008$ & $0.050 \pm 0.003$ & $0.615 \pm 0.010$ \\ \hline
    \color{blue} $^{222}$Rn & 5590.3 & $0.948 \pm 0.011$ & $0.089 \pm 0.003$ & $0.930 \pm 0.012$ \\ \hline
    \color{darkgreen} $^{224}$Ra & 5788.9 & $0.561 \pm 0.008$ & $0.055 \pm 0.003$  & $0.615 \pm 0.010$ \\ \hline
    \color{red} $^{223}$Ra & 5979.3 & $1.342 \pm 0.013$ & $0.352 \pm 0.007$ & $1.422 \pm 0.015$ \\ \hline
    \color{blue} $^{218}$Po & 6114.7 & \multirow{3}{*}{$2.811 \pm 0.017$} & \multirow{3}{*}{$0.458 \pm 0.007$} & \multirow{2}{*}{$2.409 \pm 0.019$} \\ \cline{1-2}
    \color{red} $^{227}$Th & 6146.4 & & & \\ \cline{1-2} \cline{5-5}
    \color{darkgreen} $^{212}$Bi & 6207.1 & & & $0.561 \pm 0.014$ \\ \hline
    \color{darkgreen} $^{220}$Rn & 6404.7 & $0.713 \pm 0.016$  & $0.068 \pm 0.006$ & $0.741 \pm 0.019$ \\ \hline
    \color{red} $^{211}$Bi & 6750.5 & $1.416 \pm 0.013$  & $0.324 \pm 0.006$ & $1.505 \pm 0.015$ \\ \hline
    \color{darkgreen} $^{216}$Po & 6906.5 & $0.465 \pm 0.021$ & $ 0.044 \pm 0.007$ & $0.510 \pm 0.024$ \\ \hline

  \end{tabular}
  \caption{Activities of single $\alpha$-lines for the three crystals Rita, Daisy and Verena (energy values taken from literature \cite{Firestone}). The different decay chains are marked by the colors blue ($^{238}$U), red ($^{235}$U) and green ($^{232}$Th). If two or more lines could not clearly be disentangled, the activity extends over the sum of several isotopes. Where no lines could be seen for a certain isotope, an upper limit was estimated from the events in the energy region expected. The table only includes statistical errors. The systematic error can be estimated to be about 5\,\%~-~\,10\,\% by using the activity deviations between isotopes that should be in equilibrium as, e.g., $^{223}$Ra and $^{211}$Bi. 
  }
  \label{tab:SingleAlphaAct}
\end{table}
Table~\ref{tab:SingleAlphaAct} lists all single $\alpha$-activities determined for these three detectors including the expected energy deposition in the crystal.
Almost all isotopes were found to be intrinsic contaminations since the deposited energy is equal to the Q$_{\alpha}$-value meaning that both - $\alpha$-particle and daughter nucleus - are detected simultaneously. 
If, however, an $\alpha$-emitter is located on the surface of the crystal or of any surrounding material it can happen that only the $\alpha$-particle deposits its energy in the crystal whereas the daughter nucleus is stopped outside the crystal.
This was found to be the case for $^{210}$Po.
In case two or more individual peaks in the spectrum could not clearly be disentangled, the activity in table~\ref{tab:SingleAlphaAct} is given as a sum of several isotopes.
Where no line could be seen for a certain isotope an upper limit was estimated from the number of events in the energy region expected.
\newline
Mainly decays of the natural decay chains but also a few rare earth elements ($^{144}$Nd, $^{152}$Gd, $^{147}$Sm) as well as $^{180}$W contribute to the $\alpha$-contamination.
\newline
The $\alpha$-decay of $^{180}$W, which was in 2004 for the first time unambiguously detected using CRESST CaWO$_4$ detectors \cite{TungstenAlpha}, is visible in all crystals and is used to check the analysis.
Combining the $^{180}$W events of the three crystals in table~\ref{tab:SingleAlphaAct} leads to a half-life of $T_{\frac{1}{2}} = (1.59\,\pm\,0.05) \cdot 10^{18}\,\text{y}$.
This is in good agreement with the value of $T_{\frac{1}{2}} = (1.8\,\pm\,0.2) \cdot 10^{18}\,\text{y}$ calculated in reference \cite{TungstenAlpha}.
\newline
Several short-lived isotopes (compared to the measuring time) of the natural decay chains listed in table~\ref{tab:SingleAlphaAct} should be in equilibrium with their respective long-lived mother nuclei and should, thus, exhibit the same activities.
In the $^{232}$Th chain this includes the decays of $^{228}$Th, $^{224}$Ra, $^{220}$Rn, $^{216}$Po and $^{212}$Bi.
In the $^{235}$U chain the decays of the isotopes $^{227}$Ac, $^{227}$Th, $^{223}$Ra and $^{211}$Bi should be in equilibrium and in the $^{238}$U chain the decays of $^{226}$Ra, $^{222}$Rn and $^{218}$Po are concerned.
The equilibrium was found to be valid for most isotopes within statistical errors.   
For $^{227}$Ac and $^{212}$Bi the measured counting rate was divided by the branching ratio of 1.38\,\% and 36\,\%, respectively.\footnote{In the cases where the $^{227}$Ac- and $^{212}$Bi-lines are not clearly separated from the neighbouring isotopes the activity was corrected by estimating the missing count rate from the activity the isotope should have due to the equilibrium condition.}
For all other $\alpha$-decaying isotopes listed the branching ratio is nearly 100\,\%. 
Furthermore, also the tabulated activities of $^{220}$Rn and $^{216}$Po are corrected values.
As the half-life ($150\,$ms) of the $^{216}$Po decay following the decay of its mother nucleus $^{220}$Rn is in the range of the record length used in CRESST ($\sim\,328\,$ms), the detection of a single $^{220}$Rn pulse within the record window has a probability of 32.12\% only.
Pile-up events with two pulses in the same record are in most cases removed by data quality cuts.
Due to the dead time after a record and due to cuts removing events early in the record a single $^{216}$Po event can only be found when it decays at least $\sim$~425\,ms after the $^{220}$Rn event. 
\newline
An explanation of the overestimated correction of the $^{220}$Rn activity might be an overestimation of the efficiency of data quality cuts discarding pile-up events which can, however, not be quantified.
Further deviations as, e.g., between the activities of $^{223}$Ra and $^{211}$Bi (in the order of 5\,\%~-~10\,\%) are due to systematic errors that could not be calculated as, e.g., cut efficiencies are not exactly known. 
\newline
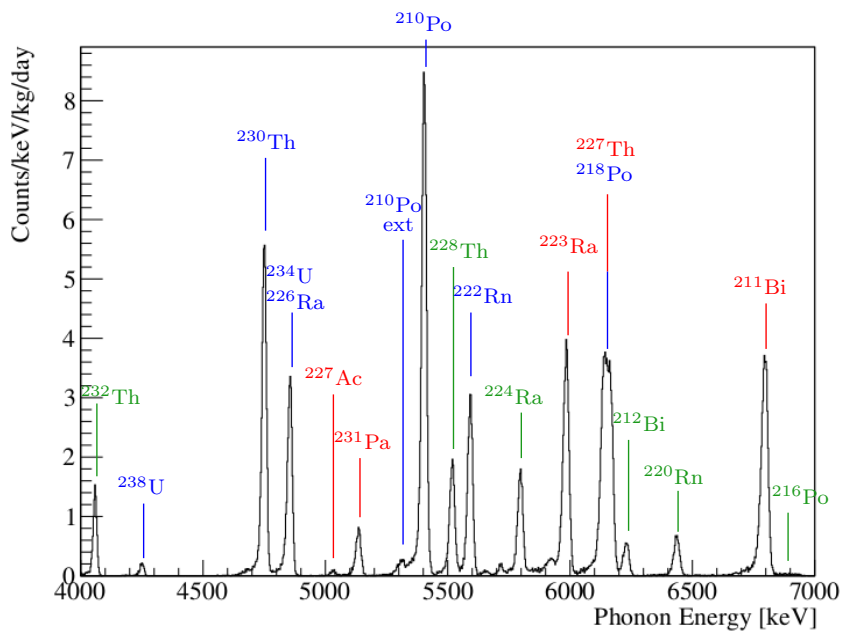
\begin{figure}[t]
  \centering
  
      \begin{overpic}[trim = 0 0 0 0, clip, width=.78\textwidth]{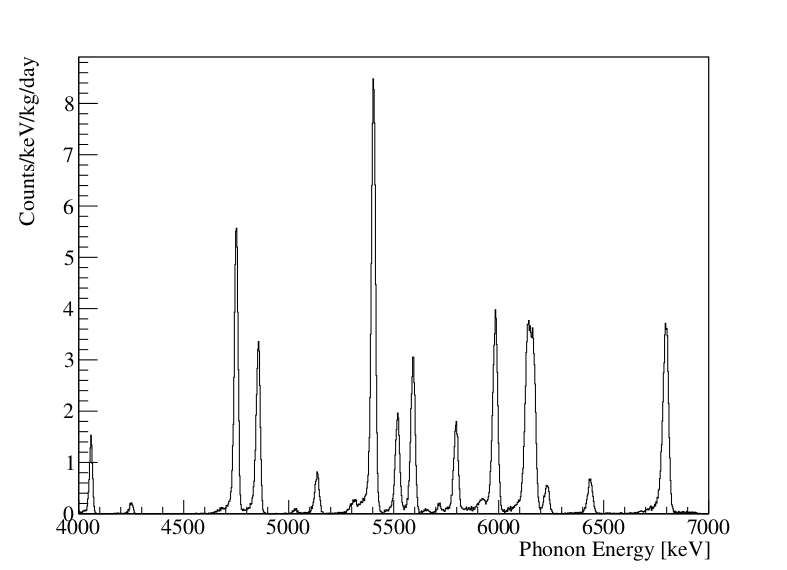}

      \put(10.7,26){\color{darkgreen} \linethickness{0.18mm} \line(0, -1){8}}
      \put(10,26){\makebox{\color {darkgreen} \scriptsize{$^{232}$Th}}}
      
      \put(15.8,15){\color{blue} \linethickness{0.18mm} \line(0, -1){6}}
      \put(14,16){\makebox{\color{blue} \scriptsize{$^{238}$U}}}

      \put(29.1,53){\color{blue} \linethickness{0.18mm} \line(0, -1){8}}
      \put(27,54){\makebox{\color {blue} \scriptsize{$^{230}$Th}}}
            
      \put(32,36){\color{blue} \linethickness{0.18mm} \line(0, -1){6}}
      \put(30.2,38){\makebox{\parbox{1.1cm}{\color{blue} \scriptsize{$^{234}$U $^{226}$Ra}}}}
      
      \put(36.5,27){\color{red} \linethickness{0.18mm} \line(0, -1){18}}
      \put(34.4,28){\makebox{\color{red} \scriptsize{$^{227}$Ac}}} 
      
      \put(39.4,20){\color{red} \linethickness{0.18mm} \line(0, -1){6.5}}
      \put(37.6,21){\makebox{\color{red} \scriptsize{$^{231}$Pa}}}
      
      \put(44.1,44){\color{blue} \linethickness{0.18mm} \line(0, -1){33.5}}
      \put(41,47){\makebox{\color{blue} \scriptsize{$^{210}$Po}}}
      \put(43,45){\makebox{\color{blue} \scriptsize{ext}}}  
      
      \put(46.6,66){\color{blue} \linethickness{0.18mm} \line(0, -1){3}}
      \put(44.3,67){\makebox{\color{blue} \scriptsize{$^{210}$Po}}}
      
      \put(49.6,41){\color{darkgreen} \linethickness{0.18mm} \line(0, -1){20}}
      \put(48,42){\makebox{\color{darkgreen} \scriptsize{$^{228}$Th}}}
      
      \put(51.5,36){\color{blue} \linethickness{0.18mm} \line(0, -1){7}}
      \put(50.6,37){\makebox{\color{blue} \scriptsize{$^{222}$Rn}}}
      
      \put(57,25){\color{darkgreen} \linethickness{0.18mm} \line(0, -1){5}}
      \put(54,26){\makebox{\color {darkgreen} \scriptsize{$^{224}$Ra}}}

      \put(62.1,40.5){\color{red} \linethickness{0.18mm} \line(0, -1){7}}
      \put(60,42.5){\makebox{\color{red} \scriptsize{$^{223}$Ra}}}
      
      \put(66.4,49){\color{red} \linethickness{0.18mm} \line(0, -1){8.5}}
      \put(66.4,40.5){\color{blue} \linethickness{0.18mm} \line(0, -1){8.5}}
      \put(64,52){\makebox{\parbox{1.1cm}{\scriptsize{\color{red}$^{227}$Th \color{blue}$^{218}$Po}}}}
      
      \put(68.7,22){\color{darkgreen} \linethickness{0.18mm} \line(0, -1){10}}
      \put(68,23){\makebox{\color{darkgreen} \scriptsize{$^{212}$Bi}}}
      
      \put(74.1,16.4){\color{darkgreen} \linethickness{0.18mm} \line(0, -1){4.5}}
      \put(71.4,17.2){\makebox{\color{darkgreen} \scriptsize{$^{220}$Rn}}}

      \put(83.7,37){\color{red} \linethickness{0.18mm} \line(0, -1){5}}
      \put(81.2,38){\makebox{\color {red} \scriptsize{$^{211}$Bi}}}
      
	  \put(86.1,14){\color{darkgreen} \linethickness{0.18mm} \line(0, -1){5.3}}
      \put(85.4,15){\makebox{\color{darkgreen} \scriptsize{$^{216}$Po}}}

    \end{overpic}
    
    \caption{$\alpha$-energy spectrum between 4\,MeV and 7\,MeV for the crystal Verena. The different peaks are labelled with the corresponding decaying nuclides of the $^{238}$U chain (blue), the $^{235}$U chain (red) and the $^{232}$Th chain (green). Two labels at the same line indicate that the two decays cannot be disentangled. Almost all decays are due to intrinsic contaminations of the crystal since the energy of the $\alpha$-particle is deposited simultaneously with the energy of the daughter nucleus (Q$_{\alpha}$-value). For $^{210}$Po a second line occurs originating from an external contamination of the surface of the crystal or of the surrounding materials. In this case the recoil energy of the daughter nuclide $^{206}$Pb ($103\,\text{keV}$) is not detected.}
    \label{fig:AlphaSpecVerena}
\end{figure}
As an example, the $\alpha$-spectrum of the crystal Verena in the energy range between 4\,MeV and $7\,\text{MeV}$ is shown in figure~\ref{fig:AlphaSpecVerena}   including all (single) $\alpha$-decays of the natural decay series of $^{238}$U (blue), $^{235}$U (red) and $^{232}$Th (green).
Two labels at the same line express that the two decays cannot be disentangled.
For $^{210}$Po a second line occurs in the spectrum due to an external contamination.
\newline
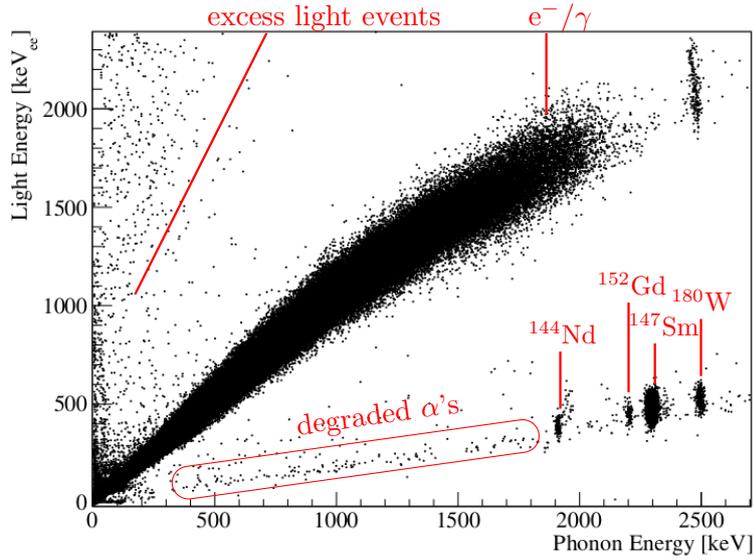
\begin{figure}[t]
  \centering
  \begin{overpic}[trim = 0 0 0 0, clip, width=.7\textwidth]{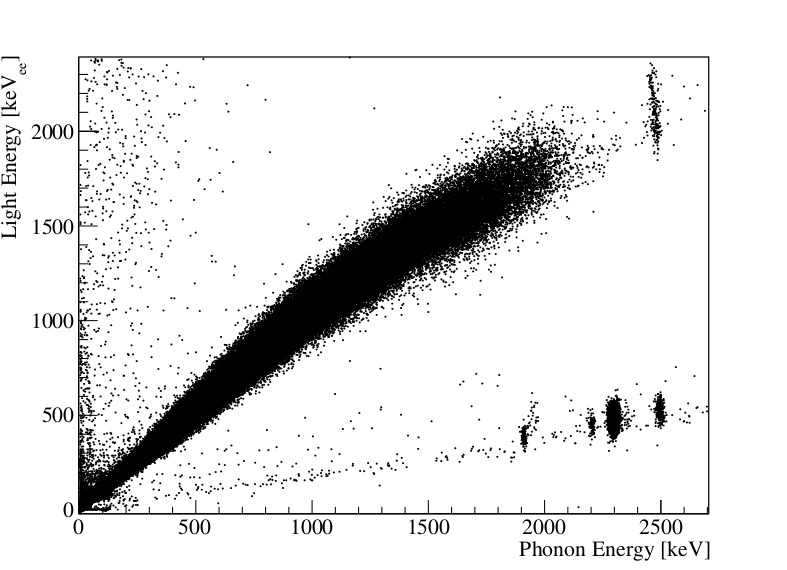}

      \put(30,65){\color{red} \linethickness{0.3mm} \line(-1, -2){16}}
      \put(24,66){\makebox{\color {red} excess light events}}
 
      \put(64,65){\color{red} \linethickness{0.3mm} \line(0, -1){10}}
      \put(63,66){\makebox{\color {red} e$^-$/$\gamma$}}
  
      \put(65.7,26){\color{red} \linethickness{0.3mm} \line(0, -1){7}}
      \put(63,27){\makebox{\color {red} \small{$^{144}$Nd}}}
      
      \put(74.0,32){\color{red} \linethickness{0.3mm} \line(0, -1){11}}
      \put(71.4,33){\makebox{\color {red} \small{$^{152}$Gd}}}
      
      \put(77.2,27){\color{red} \linethickness{0.3mm} \line(0, -1){5}}
      \put(75.2,28){\makebox{\color {red} \small{$^{147}$Sm}}}
      
      \put(82.8,30){\color{red} \linethickness{0.3mm} \line(0, -1){7}}
      \put(80.4,31){\makebox{\color {red} \small{$^{180}$W}}}
      
      \put(42,12.8){\color{red} \begin{rotate}{8.7} \oval(45, 4) \end{rotate}}
      \put(35,15.8){\makebox{\color {red} \begin{rotate}{8.7} degraded $\alpha$'s \end{rotate}}}
  
  \end{overpic}

  \caption{Light energy versus phonon energy for the low-energy $\alpha$-range of the crystal Daisy (exposure after cuts 90.10\,kg\,days). In addition to the $\alpha$-decays of the rare earth elements $^{144}$Nd, $^{152}$Gd and $^{147}$Sm as well as of $^{180}$W a continuous band of degraded $\alpha$-events is visible towards lower energies. They originate from external $\alpha$-decays where only part of the $\alpha$-energy is deposited in the crystal \cite{KSchaeffner_PHD}. Excess light events above the e$^-$/$\gamma$-band originate from a coincident detection of the crystal's scintillation light and of an additional signal in the light detector as, e.g., electrons back-scattered from the crystal into the light detector.}
  \label{fig:DaisyEE}

\end{figure}
$\alpha$-events with energies below $2.7\,\text{MeV}$ can be seen in figure~\ref{fig:DaisyEE} for the crystal Daisy.
In addition to the $\alpha$-decays of $^{180}$W and of the rare earth elements $^{147}$Sm, $^{152}$Gd and $^{144}$Nd a continuous band of degraded $\alpha$-particles is visible.
Their origin are external $\alpha$-decays within non-scintillating materials in the vicinity of the crystal where part of the energy is absorbed by this material before the $\alpha$-particle reaches the crystal \cite{KSchaeffner_PHD}.
\newline
\begin{figure}[t]
  \centering
  \begin{overpic}[trim = 0 0 0 0, clip, width=.7\textwidth]{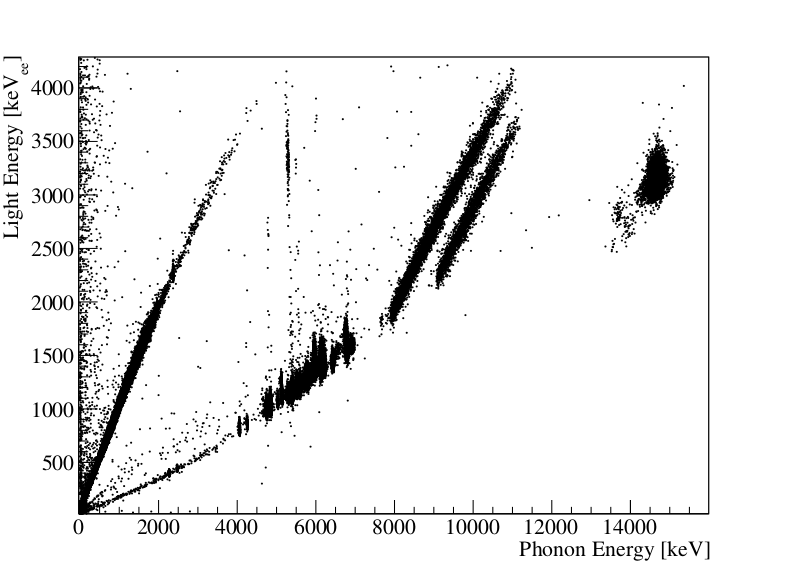}

  	  \put(59,47){\color{red} \begin{rotate}{62} \thicklines \oval(39, 10) \end{rotate}}
  	  \put(38,24){\color{red} \begin{rotate}{35} \thicklines \oval(24, 10) \end{rotate}}
  	  \put(19.5,32){\color{red} \begin{rotate}{67} \thicklines \oval(48, 6) \end{rotate}}
  	  \put(36.5,52){\color{red} \begin{rotate}{90} \thicklines \oval(15, 3) \end{rotate}}
  	  \put(82,48){\color{red} \thicklines \circle{16}}
  	  
      \put(18,58.2){\makebox{\color {red} e$^-$/$\gamma$-band}}
      \put(27,57.2){\color{red} \thicklines \line(0,-1){3}}
      \put(36,61){\makebox{\color {red} $^{210}$Po ext}}
      \put(41,60.5){\color{red} \thicklines \line(-1,-1){3}}
      \put(58,68){\makebox{\color {red} $\beta+\alpha$ cascades}}
      \put(66,67){\color{red} \thicklines \line(0,-1){2}}
      \put(72,59){\makebox{\color {red} $\alpha + \alpha$ cascades}}
      \put(82,58){\color{red} \thicklines \line(0,-1){1.7}}
      \put(30,32){\makebox{\color {red} $\alpha$-lines}}
      \put(36,31){\color{red} \thicklines \line(0,-1){2.2}}
      
      \put(79.5,34){\makebox{\color {red} \small{$^{219}$Rn/$^{215}$Po}}}
      \put(84,37){\color{red} \thicklines \line(0,1){8}}
      
      \put(71,30){\makebox{\color {red} \small{$^{220}$Rn/$^{216}$Po}}}
      \put(79,33){\color{red} \thicklines \line(0,1){9}}
      
      \put(45,18){\makebox{\color {red} \small{$^{214}$Bi/$^{214}$Po}}}
      \put(52,21){\color{red} \thicklines \line(0,1){12}}
      
      \put(53,22){\makebox{\color {red} \small{$^{212}$Bi/$^{212}$Po}}}
      \put(60,25){\color{red} \thicklines \line(0,1){17}}      
  
  \end{overpic}

  \caption{Apart from single $\alpha$-lines, cascade events turn up at higher energies as is shown here in case of the crystal Rita (exposure after cuts 92.78\,kg\,days). If the half-life of two consecutive decays is small compared to typical pulse decay times ($\sim 10\,\text{ms}-100\,\text{ms}$) the two events cannot be separated. $\beta$-$\alpha$-cascades like the $^{214}$Bi/$^{214}$Po- and $^{212}$Bi/$^{212}$Po-decays are visible in two bands shifted in parallel to the electron recoil band towards higher energies. The $\alpha$-$\alpha$ cascade decays $^{220}$Rn/$^{216}$Po and $^{219}$Rn/$^{215}$Po, on the other hand, result in two populations around $13.3\,\text{MeV}$ and $14.5\,\text{MeV}$. The line of the external $^{210}$Po $\alpha$-decay shifted towards higher light energies originates from decays where the recoiling $^{206}$Pb nucleus is directly absorbed in the light detector.}
  \label{fig:RitaCascades}

\end{figure}
At energies higher than $\sim 7\,\text{MeV}$ additional populations of events can be found (see figure~\ref{fig:RitaCascades}) due to cascade decays with half-lives shorter than the typical pulse decay times ($\sim 10\,\text{ms}-100\,\text{ms}$).
The following $\beta$-$\alpha$-cascades belong to such decays:
\begin{equation} 
\label{eq:BiPo214}
^{214}\text{Bi} \quad \xrightarrow[19.9\,\text{m}]{\beta~(3272\,\text{keV})}\quad ^{214}\text{Po} \quad \xrightarrow[164.3\,\mu\text{s}]{\alpha~(7833.5\,\text{keV})} \quad ^{210}\text{Pb}
\end{equation}
and
\begin{equation} 
\label{eq:BiPo212}
^{212}\text{Bi} \quad \xrightarrow[60.55\,\text{m}]{\beta~(2254\,\text{keV})}\quad ^{212}\text{Po} \quad \xrightarrow[0.3\,\mu\text{s}]{\alpha~(8954.1\,\text{keV})} \quad ^{208}\text{Pb}
\end{equation}
\newline
In each case, the two decays happen almost simultaneously because of the short half-life of $^{214}$Po ($^{212}$Po) of $164.3\,\mu\text{s}$ ($0.3\,\mu\text{s}$) and can, thus, not be separated by the detectors.
The spectrum of the $\beta$-decay is continuous between 0 and the endpoint energy (given in brackets in (\ref{eq:BiPo214}) and (\ref{eq:BiPo212})).
Together with the discrete $\alpha$-energy such a cascade leads to a band parallel to the e$^-$/$\gamma$-band shifted by the respective $\alpha$-energy.
In figure~\ref{fig:RitaCascades} two bands are displayed due to the two different $\beta$-$\alpha$-cascades with different starting energies.
\newline
If two $\alpha$-decays superimpose in the same way as in case of
\begin{equation} 
\label{eq:Rn219}
^{219}\text{Rn} \quad \xrightarrow[3.96\,\text{s}]{\alpha~(6946.1\,\text{keV})}\quad ^{215}\text{Po} \quad \xrightarrow[1.78\,\text{ms}]{\alpha~(7526.4\,\text{keV})} \quad ^{211}\text{Pb}
\end{equation}
and partially - as already described - of 
\begin{equation} 
\label{eq:Rn220}
^{220}\text{Rn} \quad \xrightarrow[55.6\,\text{s}]{\alpha~(6404.7\,\text{keV})}\quad ^{216}\text{Po} \quad \xrightarrow[0.15\,\text{s}]{\alpha~(6906.5\,\text{keV})} \quad ^{212}\text{Pb}
\end{equation}
this results in the two populations around $13.3\,\text{MeV}$ and $14.5\,\text{MeV}$ in figure~\ref{fig:RitaCascades}.
\newline
In addition, figure~\ref{fig:RitaCascades} also shows a line of the external $^{210}$Po $\alpha$-decay (compare table~\ref{tab:SingleAlphaAct} and figure~\ref{fig:AlphaSpecVerena}) that is shifted towards higher light energies.
This happens in case the recoiling $^{206}$Pb nucleus is absorbed in the light detector.
\newline
\begin{figure}[t]
  \centering
  \begin{overpic}[trim = 0 0 0 0, clip, width=.65\textwidth]{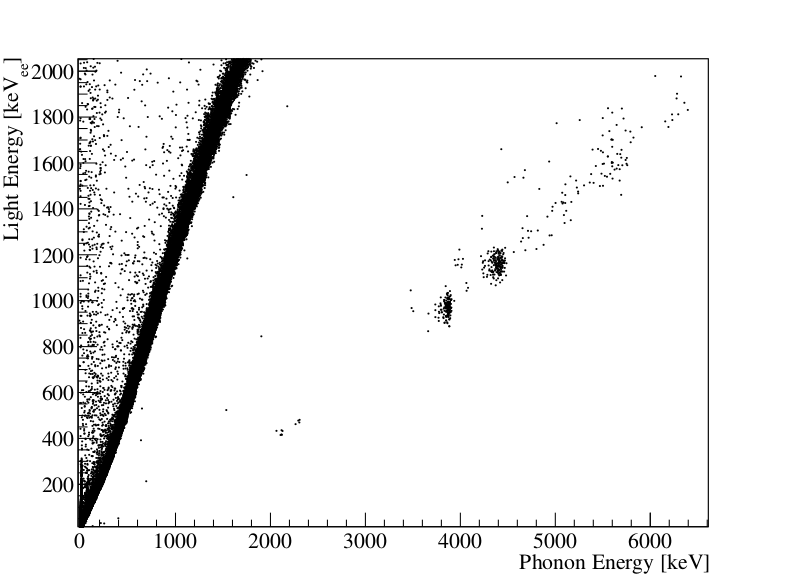}
  
  	  \put(60,42){\color{red} \begin{rotate}{38} \thicklines \oval(70, 14) \end{rotate}}
  	  \put(65,32){\makebox{\color {red} $\alpha$-band}}
      \put(69,34.5){\color{red} \thicklines \line(0,1){5.4}}

  	  \put(35,55){\makebox{\color {red} e$^-$/$\gamma$-band}}
      \put(34.5,55.7){\color{red} \thicklines \line(-1,0){6}}

  \end{overpic}

  \caption{Light energy versus phonon energy plane resulting from a measurement of the crystal TUM40 in the test setup at the LNGS (here only part of the measurement with an exposure after cuts of 1.61\,kg\,days is plotted). Due to the poor statistics the origin of the two dominating $\alpha$-lines cannot be identified which would be necessary for a precise energy calibration. The total $\alpha$-activity is competitive with the best commercial crystals with respect to radiopurity. Excess light events above the e$^-$/$\gamma$-band originate from a coincident detection of the crystal's scintillation light and of an additional signal in the light detector as, e.g., electrons back-scattered from the crystal into the light detector.}
  \label{fig:WilhelmAlphaSpec}

\end{figure}
In comparison to the commercial crystals, significantly less statistics ($\sim$20 times less live time) were achieved for the TUM crystals. 
A light energy versus phonon energy plot measured with the crystal TUM40 is shown in figure~\ref{fig:WilhelmAlphaSpec}.
Due to the poor statistics the origin of the two dominating $\alpha$-lines with activities of $1.088\,\pm\,0.089\,\text{mBq/kg}$ and $1.124\,\pm\,0.090\,\text{mBq/kg}$, respectively, cannot be identified which would be necessary for a precise energy calibration.
\newline
Even without the identification of single $\alpha$-lines it could be shown that the TUM-grown crystals are at least competitive with the commercial crystals concerning radiopurity as they were found to have the lowest total $\alpha$-activities (table~\ref{tab:AlphaAct}).
This result motivated the installation of TUM-grown crystals in the CRESST experiment. A new dark matter run of the experiment has started in summer 2013 including in addition to TUM40 also three other TUM-grown crystals (TUM29, TUM38, TUM45).
Thus, this run will also offer the possibility for a more precise investigation of their radiopurity based on much higher statistics.

\section{Summary and Conclusion}
\label{sec:conclusion}
We have investigated the radiopurity of several CaWO$_4$ crystals grown by the TUM as well as by other institutes and of the raw materials used for the production of TUM crystals.
Measurements with HPGe detectors have shown that in two samples of the CaCO$_3$ powders used for crystal growth a contamination with $^{226}$Ra of $\sim25\,\text{mBq/kg}$ and $\sim60\,\text{mBq/kg}$ depending on the supplier of the raw material is present. It was further shown that Ra is rejected during crystal growth and accumulated in the melt with an estimated segregation coefficient of $s_{\text{Ra}}<0.12$ (90\%~CL).
Such a behavior was also observed for Th. 
This offers the possibility to improve the radiopurity of the crystals in a reproducible way by using only a fresh melt for crystal growth as well as by multiple crystallization steps \cite{danevich11}. \\
The total $\alpha$-activities of TUM-grown crystals as low as $1.23\,\pm\,0.06\,\text{mBq/kg}$ were found to be significantly smaller than the activities of crystals from other institutes ranging between $3.05\,\pm\,0.02\,\text{mBq/kg}$ and $107.13\,\pm\,0.14\,\text{mBq/kg}$. 
Generally, $\alpha$-decays can easily be identified and due to their high energies are not a background for dark matter search.
However, several of them are correlated to low-energy $\beta$-decays and $\gamma$-emissions (in the same decay chains).
Such intrinsic contaminations of the crystal including, e.g., the isotope $^{227}$Ac belong to the dominating backgrounds for dark matter search.
The reduction of such backgrounds has direct impact on the sensitivity of CRESST for dark matter search and is, thus, of crucial importance.
A quantitative study is currently under investigation.
\\
A cross-check of activities determined by $\gamma$-spectrometry and by analysis of $\alpha$-decays in low-temperature detectors is planned for the future.
Several of the crystals produced at the TUM are now installed in CRESST which has started a new dark matter run in summer 2013.
This will also allow a more precise determination of the crystals' intrinsic $\alpha$-activities.
In addition, the investigation of the contamination with $\beta$-decaying isotopes will help to further improve the radiopurity of the crystals. 

\acknowledgments
This research was supported by the DFG cluster of excellence ``Origin and Structure of the Universe'', by the Helmholtz Alliance for Astroparticle Physics, by the Maier-Leibnitz-Laboratorium (Garching), by the BMBF: Project 05A11WOC EURECA-XENON, by the Spanish Ministerio de Economía y Competitividad and the European Regional Development Fund (MINECO-FEDER) (FPA2011-23749), and by the Consolider-Ingenio 2010 Programme under grant MULTIDARK CSD2009-00064. We would like to express our gratitude to the members of GIFNA group of Zaragoza University for their help with radiopurity measurements.

\bibliographystyle{JHEP}
\bibliography{BibtexDatabase}{}

\end{document}